\documentclass[preprint,draft,prl,12pt,onecolumn]{revtex4}

\input{tcilatex}

\begin{document}

\title{Local conversion of GHZ states to approximate W states}
\author{P. Walther$^{1\ast }$, K.J. Resch$^{1\ast }$, and A. Zeilinger$%
^{1,2} $}
\affiliation{$^{1}$Institut f\"{u}r Experimentalphysik, Universit\"{a}t Wien,
Boltzmanngasse 5, 1090 Vienna, Austria\\
$^{2}$Institut f\"{u}r Quantenoptik und Quanteninformation, \"{O}%
sterreichische Akademie der Wissenschaften, Austria\\
$^{\ast }$These authors contributed equally to this work.}

\begin{abstract}
Genuine 3-qubit entanglement comes in two different inconvertible types
represented by the GHZ state\ and the W state. \ We describe a specific
method based on local positive operator valued measures (POVMs) and
classical communication that can convert the ideal $N$-qubit GHZ state to a
state arbitrarily close to the ideal $N$-qubit W state. \ We then
experimentally implement this scheme in the 3-qubit case and characterize
the input and the final state using 3-photon quantum state tomography.
\end{abstract}

\maketitle

Entanglement is at the heart of quantum mysteries and the power of quantum
information. \ While 2-qubit entanglement is fairly well understood with
good entanglement measures \cite{entmeasures}, understanding multi-qubit
entanglement remains a considerable challenge. \ With 3-qubits it is no
longer enough simply to ask \emph{if} the qubits are entangled or not; one
must also ask \emph{how} the qubits are entangled. \ The two most common
examples of 3-qubit entangled states are the $\left| \mathrm{GHZ}%
\right\rangle =1/\sqrt{2}\left( \left| HHH\right\rangle +\left|
VVV\right\rangle \right) $ \cite{ghz} and $\left| \mathrm{W}\right\rangle =1/%
\sqrt{3}\left( \left| HVV\right\rangle +\left| VHV\right\rangle +\left|
VVH\right\rangle \right) $ \cite{ZHG,cirac}, where $\left| H\right\rangle $
and $\left| V\right\rangle $ represent horizontally- and vertically-
polarized photon states. \ An important characteristic of 3-particle GHZ\
states is that loss of any one of the qubits leaves the other two in a mixed
state with only classical correlations. \ \ It is well-known that if two
qubits are maximally entangled, then neither can be entangled to a third;
the W state is the 3-qubit state in which each pair of qubits have the same
and maximum amount of bipartite entanglement. \ This feature makes the
entanglement of the W state maximally symmetrically robust against loss of
any single qubit. \ It has been shown that the states $\left| \mathrm{GHZ}%
\right\rangle $ and $\left| \mathrm{W}\right\rangle $ represent two distinct
classes of 3-qubit entanglement that cannot be interconverted under any
local operations and classical communication (LOCC)\cite{cirac,acin}. \
Experimental realizations of GHZ states and more recently W states have been
performed in optical and trapped ion experiments \cite%
{dik,GHZexp,4ghz,Wexp,blatt}. \ 

While conversion of a GHZ\ state to an exact W state is not possible via
LOCC, a specific scheme based on partial quantum measurement (positive
operator valued measures or POVMs) and classical communication can, however,
convert a GHZ state to an \emph{arbitrarily good approximation} to a W state
with a tradeoff between the fidelity of the final state and the probability
of success. \ In any real experiment, where there is inevitable noise in
state production and measurement, arbitrarily good approximations are
indistinguishable from directly-prepared W states. \ In the present work, we
first discuss a POVM scheme, inspired by the procrustean method \cite%
{procrustean}, for converting the 3-qubit state $\left| \mathrm{GHZ}%
\right\rangle $ into an approximate $\left| \mathrm{W}\right\rangle $. \
Then we generalize the scheme to convert between $N$-qubit analogues of the
GHZ and W states.\ \ Finally, we experimentally apply the POVM\ scheme to a
3-photon GHZ\ state and characterize the change from the input to the output
using quantum state tomography \cite{tomography}.

In the diagonal basis, where $\left| D\right\rangle =1/\sqrt{2}\left( \left|
H\right\rangle +\left| V\right\rangle \right) $ and $\left| A\right\rangle
=1/\sqrt{2}\left( \left| H\right\rangle -\left| V\right\rangle \right) $, we
can rewrite $\left| \mathrm{GHZ}\right\rangle =1/2\left( \left|
DDD\right\rangle +\left| DAA\right\rangle +\left| ADA\right\rangle +\left|
AAD\right\rangle \right) $. \ It is apparent in this basis that $\left| 
\mathrm{GHZ}\right\rangle $ is a superposition of an unwanted term, $\left|
DDD\right\rangle ,$ and a W-state. We define a local POVM with elements $%
\varepsilon _{1}=\left| A\right\rangle \left\langle A\right| +a^{2}\left|
D\right\rangle \left\langle D\right| $ and $\varepsilon _{2}=(1-a^{2})\left|
D\right\rangle \left\langle D\right| $, where $a$ is a real number between 0
(perfect measurement) and 1 (no measurement). \ This POVM\ is applied to
each photon in the state and if all of the parties find element $\varepsilon
_{1}$, then the new state is $\left| \psi \right\rangle =\mathcal{N}\left[
a^{3}/2\left| DDD\right\rangle +a\sqrt{3}/2\left| \mathrm{W}^{\prime
}\right\rangle \right] ,$with the normalization constant, $\mathcal{N}=2/%
\sqrt{a^{6}+3a^{2}}.$ The state $\left| \mathrm{W}^{\prime }\right\rangle =1/%
\sqrt{3}\left( \left| DAA\right\rangle +\left| ADA\right\rangle +\left|
AAD\right\rangle \right) $ is simply related to $\left| \mathrm{W}%
\right\rangle $ by three single-qubit rotations. \ It is clear from this
state, that the unwanted term $\left| DDD\right\rangle $ is reduced relative
to the term $\left| \mathrm{W}^{\prime }\right\rangle $. It is also clear
that one achieves a pure $\left| \mathrm{W}^{\prime }\right\rangle $ only in
the limit as $a\rightarrow 0$ where the probability of success also goes to
zero. \ Nevertheless, the fidelity of this state with the desired $\left| 
\mathrm{W}^{\prime }\right\rangle $ is $\mathcal{F}_{\mathrm{W}^{\prime
}}=\left| \left\langle \psi |\mathrm{W}^{\prime }\right\rangle \right|
^{2}=3/(a^{4}+3)$, which rapidly rises from 3/4 to 1 as $a$ decreases, i.e.,
as the strength of the measurement increases. \ Conversely, the fidelity
with the GHZ state $\mathcal{F}_{\mathrm{GHZ}}=\left| \left\langle \psi |%
\mathrm{GHZ}\right\rangle \right| ^{2}=(a^{4}+6a^{2}+9)/(4a^{4}+12)$ drops
from 1 to 3/4 as $a$ decreases.

Our method can be generalized to convert an $N$-qubit GHZ state, $\left|
N+\right\rangle ,$ where $\left| N\pm \right\rangle =1/\sqrt{2}\left( \left|
H\right\rangle ^{\otimes N}\pm \left| V\right\rangle ^{\otimes N}\right) $,
into an arbitrarily good approximation to the $N$-qubit W state $\left| 
\mathrm{W}_{N}^{\prime }\right\rangle =1/\sqrt{N}\left( \left|
DA...A\right\rangle +\left| AD...A\right\rangle +...+\left|
AA...D\right\rangle \right) $ \cite{generalW}. \ We use the fact that the
GHZ states $\left| N\pm \right\rangle $ satisfy the following relation:%
\begin{equation}
\left| N\pm \right\rangle =\frac{1}{\sqrt{2}}\left[ \left| \left( N\text{-}%
M\right) +\right\rangle \left| M\pm \right\rangle +\left| \left( N\text{-}%
M\right) -\right\rangle \left| M\mp \right\rangle \right] ,
\end{equation}%
where $M<N$. \ Notice that this factorization preserves the evenness or
oddness in the number of negative signs. \ Through repeated application of
these two rules, one can factor $\left| N+\right\rangle $ in terms of only
single-qubit states $\left| D\right\rangle \equiv \left| 1+\right\rangle $
and $\left| A\right\rangle \equiv \left| 1-\right\rangle $. \ This
reexpresses the GHZ state as an equally weighted superposition of \emph{all} 
$2^{N-1}$ terms with an \emph{even number} of $\left| A\right\rangle $s. \
When $N$ is odd, the GHZ state can be directly rewritten as,%
\begin{equation}
\left| N+\right\rangle =\frac{1}{\sqrt{2^{N-1}}}\left[ \sqrt{N}\left| 
\mathrm{W}_{N}^{\prime }\right\rangle +\sqrt{2^{N-1}-N}\left| \phi
\right\rangle \right] ,  \label{genghz}
\end{equation}%
where the state $\left| \phi \right\rangle $ is a superposition of all those
terms containing an odd number and at least 3 $\left| D\right\rangle $s. \
When $N$ is even, application of a local transformation $\left|
D\right\rangle \rightarrow \left| A\right\rangle ,$ $\left| A\right\rangle
\rightarrow \left| D\right\rangle $ to any qubit allows the GHZ state to be
written in the form of Eq. \ref{genghz}. \ Applying the same local POVM\ as
in the 3-qubit case on each of the $N$ qubits, and given that each POVM\
returns element $\varepsilon _{1}$ the unwanted amplitudes by \emph{at least}
a factor of $a^{3}$ while reducing the desired amplitude by only a single
factor of $a$. \ In general, the fidelity of the resultant state by this
prescription with $\left| \mathrm{W}_{N}^{\prime }\right\rangle $ is given by%
\[
\mathcal{F}_{\mathrm{W}^{\prime }}^{N}=\frac{2a^{2}N}{\left( 1+a^{2}\right)
^{N}-\left( 1-a^{2}\right) ^{N}}
\]%
regardless of whether $N$ is even or odd.

The details on our experimental method for creating 3-photon GHZ states can
be found in \cite{ghztomo}. \ Ultraviolet laser pulses from a
frequency-doubled Ti:Sapphire laser make two passes through a type-II
phase-matched $\beta $-barium borate (BBO) crystal aligned, with walk-off
compensation to produce 2-photon pairs each in the Bell state $\left| \phi
^{+}\right\rangle $ \cite{kwiatdownconv}. \ These 2 independent photon pairs
can be further entangled when combined at the polarizing beamsplitter (PBS1)
and the four photons take four separate outputs A and B. \ Recall that a PBS
works by reflecting horizontally-polarized light $\left| H\right\rangle $
and transmitting vertically-polarized light $\left| V\right\rangle $. \
Thus, two photons that were incident from different sides can only pass into
different output modes when their polarizations were both $\left|
H\right\rangle $ or both $\left| V\right\rangle $. \ In this sense, the PBS
acts as a quantum parity check \cite{paritycheck}. \ Given that the parity
check succeeds on the two photons from the independent pairs, our state is
transformed from the product state $\left| \phi ^{+}\right\rangle
_{12}\left| \phi ^{+}\right\rangle _{34}$ to the 4-photon GHZ state $\left|
4+\right\rangle =1/\sqrt{2}\left( \left| HHHH\right\rangle _{AB14}+\left|
VVVV\right\rangle _{AB14}\right) $ \cite{4ghz}. \ We project photon 4 onto
the state $\left| D\right\rangle $ and when this projection succeeds leaves $%
\left| \mathrm{GHZ}\right\rangle =1/\sqrt{2}\left( \left| HHH\right\rangle
_{AB1}+\left| VVV\right\rangle _{AB1}\right) $.

A tomographically complete set of measurements for a 3-photon polarization
state requires 64 polarization measurements. \ We use the 64 combinations of
the single-photon projections $\left| H\right\rangle ,$ $\left|
V\right\rangle ,$ $\left| D\right\rangle ,$ and $\left| R\right\rangle $ on
each of the 3 photons. \ These projections are implemented using a
quarter-wave plate and polarizer for each of photons A and B, and a half- or
quarter-wave plate and PBS2 for photon 1. \ Successful projections are
signalled by four-photon coincidence measurements, 3 photons for the state
and 1 trigger photon, using single-photon counting APDs and coincidence
logic. \ The most-likely physical density matrix for our 3-qubits is
extracted using maximum-likelihood reconstruction \cite{maxlike,james}.

We begin with the GHZ\ state that was characterized previously via 3-photon
quantum state tomography \cite{ghztomo}. We rewrite the density matrix in
the $\left| D/A\right\rangle $ basis; this gives the density matrix shown in
Fig. 2b (real part) and 2c (imaginary part). \ A comparison to the ideal
GHZ\ written in the same basis is shown in Fig. 2a (real part only, the
imaginary part is all zero). \ The colour plots display the absolute value
of each element and show that the two matrices have the same structure.

Our POVM was implemented using three partial polarizers. \ Instead of
orienting the polarizers in the $\left| D/A\right\rangle $ basis, we rotated
the polarization of each photon by $45%
{{}^\circ}%
$ using half-wave plates. These rotations were accomplished by using the
existing half-wave plate in mode 1, and by adding two additional half-wave
plates in modes A and B. These extra rotations allowed the partial
polarizers to operate in the $\left| H/V\right\rangle $ basis, and therefore
our POVM also operates in the $\left| H/V\right\rangle $ basis. \ Each
polarizer comprised of two uncoated glass microscope slides such that the
angle of incidence for the input light was at 56$^{\circ }$, near Brewster's
angle (Figure 1). \ The configuration of the plates are such that the beam
experienced minimal additional transverse shift and maintained high coupling
efficiency into single-mode fibres. \ Such partial polarizers have been used
to study hidden nonlocality and entanglement concentration of
maximally-entangled mixed states \cite{partpols}. \ We placed each such
element so that vertically-polarized light was $P$-polarized and
horizontally-polarized light was $S$-polarized; we measured 88\%
transmission for the vertically-polarized light and 33\% for the
horizontally-polarized light. \ The transmission of the vertical light is
thus only 38\% of that for the horizontal and we can can describe the
experiment using the POVM\ elements $\varepsilon _{1}$ and $\varepsilon _{2}$
with $a^{2}\approx 38\%$. \ With this attenuation value, and beginning with
the ideal GHZ\ state, the fidelity of the state with the ideal W state given
3 POVM outcomes $\varepsilon _{1}$ is expected to increase from 75\% to
95\%. \ 

We used the same 64 tomographic measurement settings for the W state as for
the GHZ state. \ Data for each setting was accumulated for 1800 seconds and
yielded a maximum of 120 four-fold coincidence counts (for the $|VVV\rangle $
projection). \ To account for laser power drift, which was small but not
insignificant, we divided the four-folds by the square of the singles at the
trigger detector. \ Background four-folds from a two-fold coincidence count
and an uncorrelated accidental were estimated for each measurement setting\
and subtracted from the measured coincidences. \ Using the maximum
likelihood reconstruction, our most likely density matrix is shown in Fig.
3b (real part) and 3c\ (imaginary part). \ The ideal W state density matrix
consists of only 9 real elements -- 3 diagonals of 1/3 height corresponding
to $\left| HVV\right\rangle $, $\left| VHV\right\rangle $, and $\left|
VVH\right\rangle $ and 6 maximal positive coherences between them. \ It is
clear from the data that the dominant elements in the density matrix are
those same 9 elements. \ In Fig. 3a., we show the effect of the POVM\ with
our experimentally measured attenuation on the ideal GHZ state. \ The
diagonal elements are attenuated much more strongly, by $a^{2}$, while the
coherences remain maximal and are thus reduced only by $a$. \ Note that one
specific unwanted term contained in the diagonal element for $\left|
VVV\right\rangle $ is much more significant after the application of the
POVM; this noise contribution, in the ideal case, is untouched by the POVM,
and in the real case, least reduced.\ 

We characterize the changes in our states using fidelity. \ The fidelity of
a density matrix, $\rho $, with a pure quantum state, $\left| \psi
\right\rangle $, is given by $\mathcal{F}=\left\langle \psi \right| \rho
\left| \psi \right\rangle $. \ We calculate this fidelity with a general GHZ
(W)\ state, $\left| \mathrm{GHZ}_{G}\right\rangle $ $(\left| \mathrm{W}%
_{G}\right\rangle )$, which is related to $\left| \mathrm{GHZ}\right\rangle $
$(\left| \mathrm{W}\right\rangle )$ by 3 local unitary rotations. \ The
initial state has a fidelity of $\mathcal{F}_{\mathrm{GHZ}_{G}}=\left(
79.4\pm 1.6\right) \%$ and $\mathcal{F}_{\mathrm{W}_{G}}=\left( 60.5\pm
1.9\right) \%$ as compared with the ideal $100\%$ and $75\%$. \ After
successful application of three local POVMs, $\mathcal{F}_{\mathrm{GHZ}%
_{G}}=\left( 59.8\pm 2.5\right) \%$ and $\mathcal{F}_{\mathrm{W}_{G}}=\left(
68.4\pm 2.4\right) \%$. \ Uncertainties in quantities extracted from these
density matrices were calculated using a Monte Carlo routine and assumed
Poissonian errors. \ A theoretical calculation based on our measured initial
state and measured $a$ has yields a final state with $\mathcal{F}_{\mathrm{W}%
_{G}}=75\%.$ \ Thus much of the difference with the expected fidelity in the
ideal case is a result of the quality of the initial state. Nevertheless,
the overlap with a W state has been significantly improved while the overlap
with a GHZ state has been strongly reduced.

We have described a method for converting $N$-qubit GHZ\ states to
arbitrarily good approximations to $N$-qubit W states based on generalized
quantum measurements (POVMs). \ We have implemented this scheme for the
3-qubit case and characterized the input and output states using multiphoton
quantum state tomography. \ We have quantitatively shown that the
transformation induced by the partial polarizers results in a decrease in
the overlap of the state with a GHZ\ state while increasing the overlap with
the desired W state. \ Multiparticle entanglement is essential to the
success of quantum information processing. \ The theory and experimental
work presented here extends our abilities to manipulate and understand the
relationship between different types of complex entangled states.

The authors thank Antonio A\'{c}in, \v{C}aslav Brukner, Klaus Hornberger,
Morgan Mitchell, and Andrew White for valuable discussions. \ This work was
supported by ARC Seibersdorf Research GmbH, the Austrian Science Foundation
(FWF), project number SFB 015 P06, NSERC, and the European Commission,
contract number IST-2001-38864 (RAMBOQ).

\textbf{Figure 1. Experimental setup for production of GHZ\ state and its
conversion to an approximate W state. \ A double-pass pulsed parametric
down-conversion crystal (BBO) is used to create two pairs of
polarization-entangled photons both in the state }$\left| \phi
^{+}\right\rangle $\textbf{. \ Extra crystals (COMP) compensate for walkoff
effects. \ A polarizing beamsplitter (PBS) performs a parity check on two of
the photons, one from each pair; given that the parity check succeeds,
signalled by the two photons taking different output modes, the emerging
photons are in a four photon GHZ state. \ Projection of photon 4 onto the
state }$\left| D\right\rangle $\textbf{\ leaves the remaining 3 photons in
the desired 3-photon GHZ state. \ Each photon in the GHZ state was rotated
locally -- photons A and B were rotated using additional half-wave plates
(HWP) and photon 1 was rotated using the existing half-wave plate, rotating
the polarization only }$45^{\circ }$\textbf{\ instead of }$90^{\circ }$%
\textbf{. \ Our 3-qubit local POVM\ is performed using 3 partial polarizers
(PP). \ Each partial polarizer consists of two microscope slides (MS)
mounted such that the light is incident at }$56^{\circ },$ \textbf{%
Brewster's angle for }$n=1.5$\textbf{. This configuration had a\ measured
transmission of 88\% for p-polarization and 33\% for s-polarization. \ Each
POVM was oriented to reduce the horizontal component of the light relative
to the vertical component. \ The three-photon polarization measurements for
tomography were taken\ using quarter-wave plates\ (QWP) and rotatable
polarizers (POL) for photons A and B, and a half- or quarter-wave plate and
a fixed polarizer (actually a second PBS) for photon A before
photon-counting detectors (DET and TRIG). \ A fixed QWP is mode A was used
to compensate birefringence in the PBS.}

\textbf{Figure 2. Density matrix of an ideal and experimentally measured
3-photon GHZ state. \ The reconstructed density matrix of an a) ideal GHZ
state (real part only) and our measured GHZ\ state, real part b) \&
imaginary part c), from reference \cite{ghztomo}. \ The state is displayed
in the }$D/A$\textbf{-basis, where }$D=\frac{1}{\sqrt{2}}\left( \left|
H\right\rangle +\left| V\right\rangle \right) $\textbf{\ and }$A=\frac{1}{%
\sqrt{2}}\left( \left| H\right\rangle -\left| V\right\rangle \right) $ 
\textbf{are defined as in the text. \ The false colour plots display the
absolute value of each component of the matrices and are meant to show the
structure of the matrix - namely that our GHZ\ state is characterized in
this basis by 4 diagonal elements of equal height with maximal positive
coherences. \ The experimentally measured density matrix has a fidelity of
77\% with the ideal GHZ state and 79\% with any state related to the ideal
GHZ\ state via local unitary transformations.}

\textbf{Figure 3. Density matrix of approximate W output state after the
POVM procedure. \ The reconstructed density matrix of our output state, b)
real part \& c)\ imaginary part, after the three local POVM operations. \
The application of the POVMs have suppressed several of the matrix
components such that the final state contains only 9 major elements. \ These
are the same 9 elements for the ideal W state. \ The operation has increased
the fidelity of our state with a W state from 61\% to 68\% while at the same
time reducing the fidelity of our state with a GHZ\ state from 79\% to 60\%.
For comparison, we show a) the action of the POVM\ operation on the ideal
GHZ\ state. \ Although this state still contains large coherences with the
HHH component, this state has 95\% fidelity with the W state. \ }

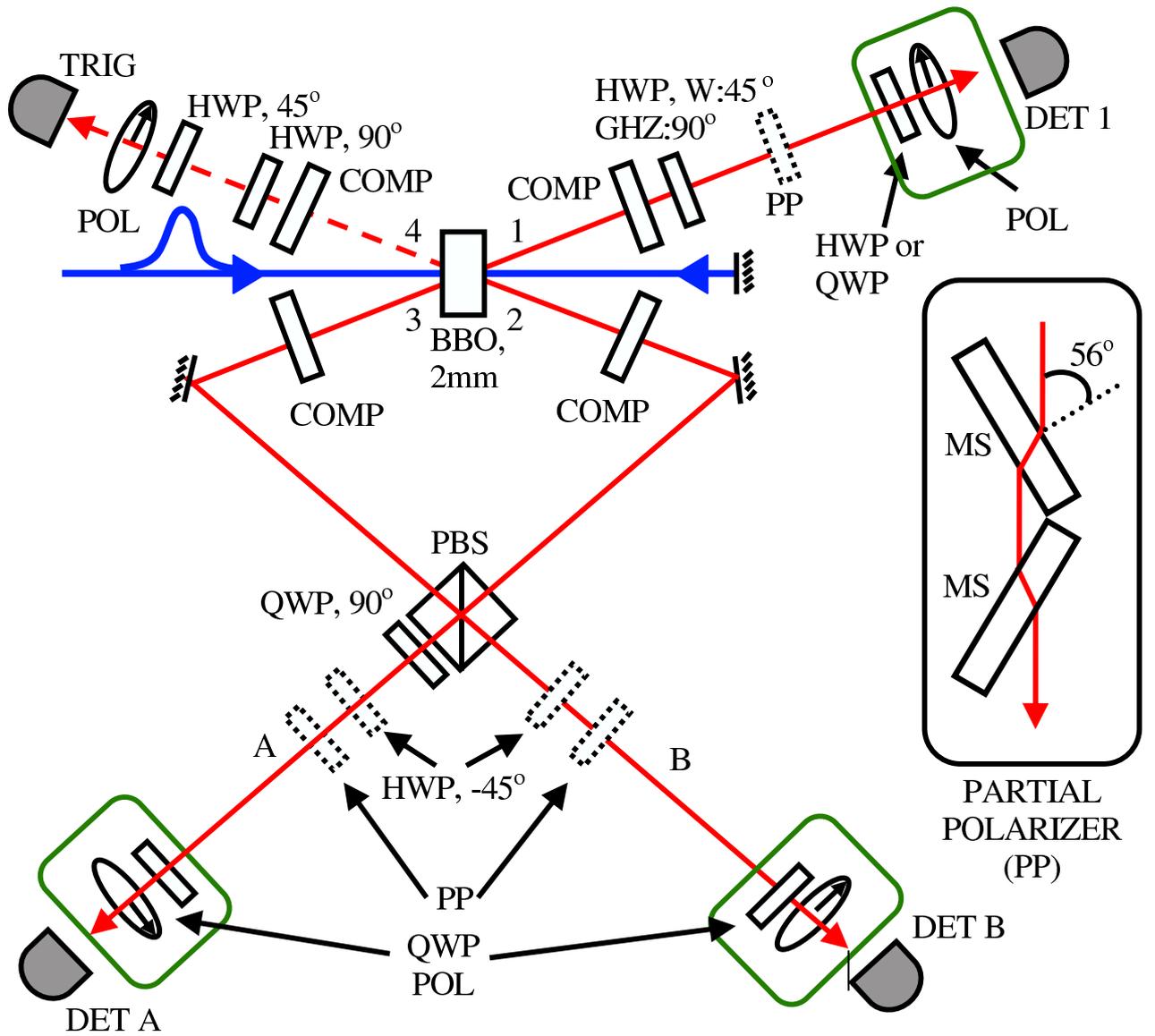
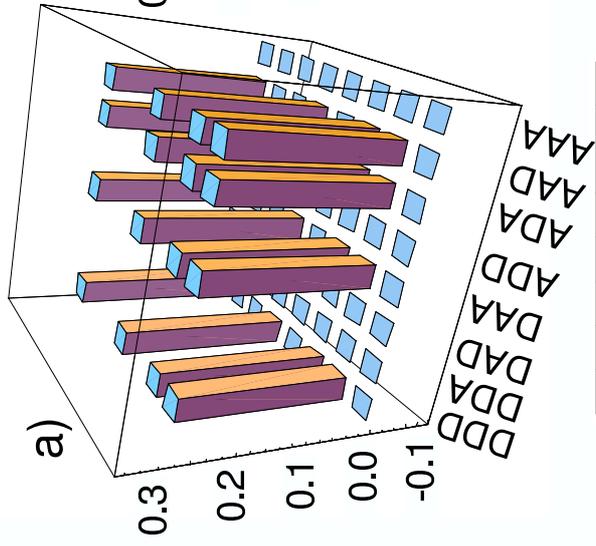
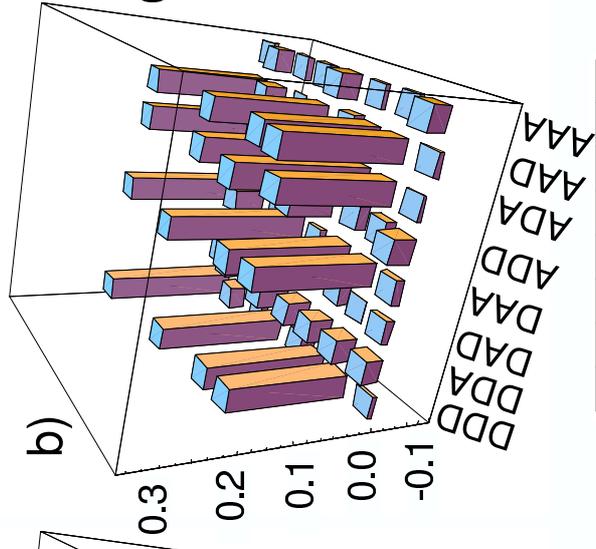
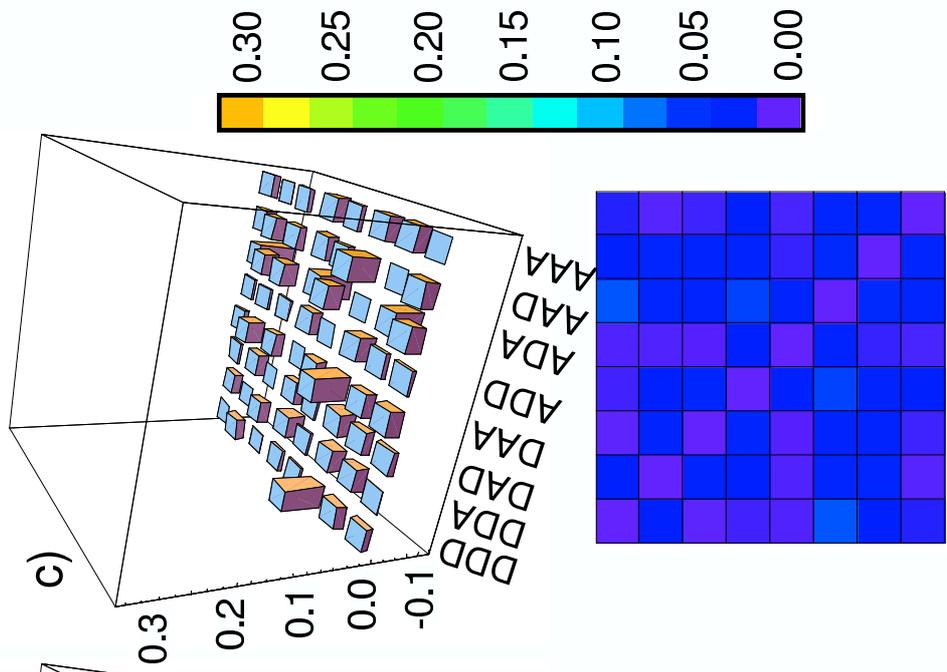
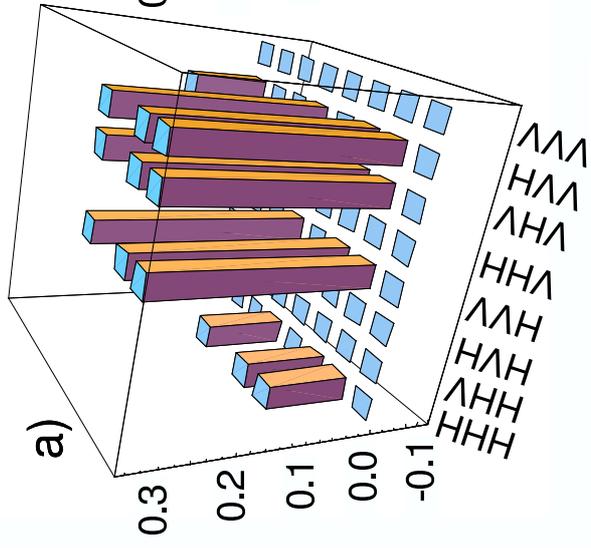
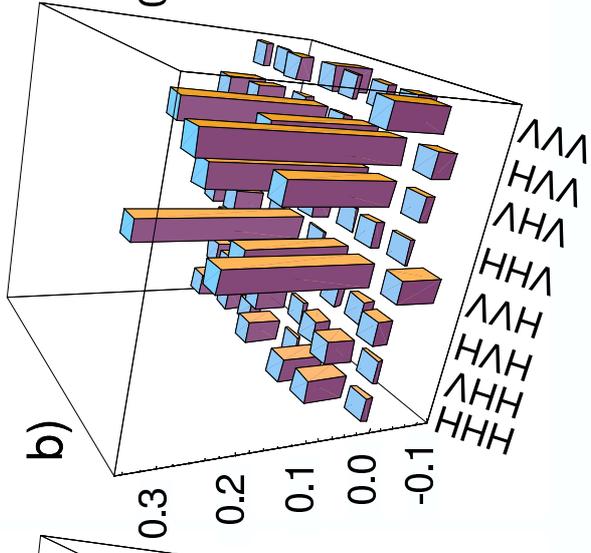
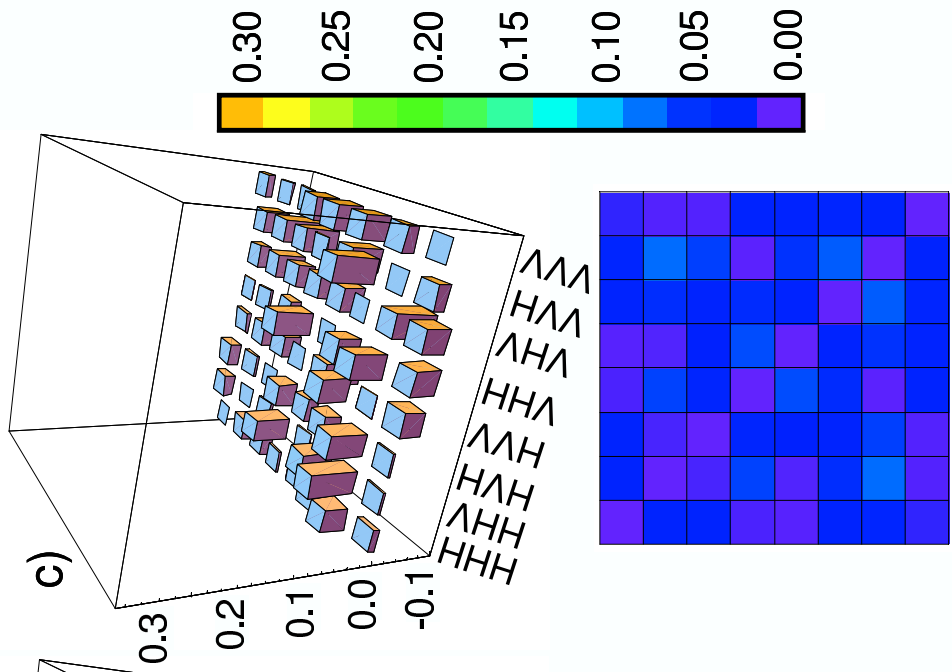
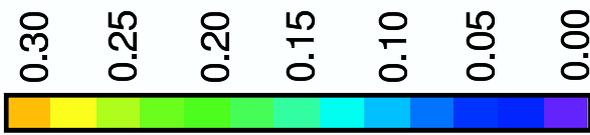
\end{document}